\documentclass[fleqn,10pt]{wlscirep}
\usepackage{bbold}

\title{Advanced-Retarded Differential Equations in Quantum Photonic Systems}

\author[1,*]{Unai Alvarez-Rodriguez}
\author[2,3]{Armando Perez-Leija}
\author[4]{I\~nigo L. Egusquiza}
\author[2]{Markus Gr\"afe}
\author[1]{Mikel Sanz}
\author[1]{Lucas Lamata}
\author[2,5]{Alexander Szameit}
\author[1,6]{Enrique Solano}

\affil[1]{Department of Physical Chemistry, University of the Basque Country UPV/EHU, Apartado 644, 48080 Bilbao, Spain}
\affil[2]{Institute of Applied Physics, Abbe Center of Photonics, Friedrich-Schiller-Universit\"{a}t Jena, Max-Wien-Platz 1, Jena 07743, Germany}
\affil[3]{Max Born Institute, Max Born Strasse 2A, 12489 Berlin, Germany}
\affil[4]{Department of Theoretical Physics and History of Science, University of the Basque Country UPV/EHU, Apartado 644, 48080 Bilbao, Spain}
\affil[5]{Institute for Physics, University of Rostock,  Albert-Einstein-Stra{\ss}e 23, 18059 Rostock, Germany}
\affil[6]{IKERBASQUE, Basque Foundation for Science, Maria Diaz de Haro 3, 48013 Bilbao, Spain}

\affil[*]{Corresponding author: unaialvarezr@gmail.com}

\begin{abstract}
We propose the realization of photonic circuits whose dynamics is governed by advanced-retarded differential equations. Beyond their mathematical interest, these photonic configurations enable the implementation of quantum feedback and feedforward without requiring any intermediate measurement. We show how this protocol can be applied to implement interesting delay effects in the quantum regime, as well as in the classical limit. Our results elucidate the potential of the protocol as a promising route towards integrated quantum control systems on a chip.   
\end{abstract}

\begin{document}

\flushbottom
\maketitle
\thispagestyle{empty}

\section*{Introduction}
In advanced-retarded (A-R) differential equations, or mixed functional differential equations, the derivative of the associated function explicitly depends on itself evaluated at different advanced-retarded values of the variable \cite{mfde1,mfde2,mfde3,mfde4,mfde5}.  In order to solve such A-R equations either analytically or numerically, we require the knowledge of the solution history out of the domain of the equation. In many scientific disciplines A-R differential equations are used to describe phenomena containing feedback and feedforward interactions in their evolution \cite{aero,eco,ner}. In physics, for instance, A-R equations can be used to model dynamical systems exhibiting  certain symmetry in the evolution. As a prominent examples one may mention the application of A-R equations \cite{wefe1} in Wheeler-Feynman absorber theory \cite{wefe2,wefe3} and in the propagation of waves in discrete spatial systems \cite{mapa}. 

In the context of Quantum Mechanics, the implementation of feedback is more intricate than in the classical case due to the sensitivity of quantum systems to measurements. In this regard, a set of techniques has been developed for the realization of feedback-dependent systems, each of them employing different resources such as dynamical delays \cite{grim,whalen,fede}, machine learning optimization \cite{realt}, weak measurements \cite{wk1,wk2}, including quantum memristors \cite{qm1,qm2}, and projective measurements for digital feedback \cite{l1}, among others. Certainly, the inclusion of feedback or memory effects in quantum dynamical systems has extended the scope of quantum protocols, and it has allowed for the study and reproduction of more complex phenomena. Therefore, devising schemes for engineering Hamiltonians that display advanced-retarded dynamics is of great relevance. Along these lines, the field of non-Markovian quantum dynamics focuses on the study of effective equations that govern the evolution of systems interacting with environments depending on previous times \cite{nmk1,nmk2,nmk3,nmk4}. As a result, estimation of non-Markovianity sheds light on the memory content of the systems under study.

In this article we show that photonic lattices can be used to effectively tailor the dynamics of classical and quantum light fields in an advanced-retarded fashion. Our strategy is to exploit the duality between light propagation in space and time evolution \cite{toni}. Our A-R photonic approach exploits the isomorphism existing between the steady state of judiciously-designed photonic waveguide circuits and solutions of A-R differential equations. We foresee that the inherent versatility of the proposed system will make the implementation of feedback and feedforward noticeably simple, in both quantum and classical frameworks, and thus may pave the way to interesting applications in integrated quantum technologies. 

\section*{Results}

\subsection*{Advanced-Retarded Analogy}
\label{sec:ara}

\begin{figure}[htbp]
\includegraphics[width=0.5\linewidth]{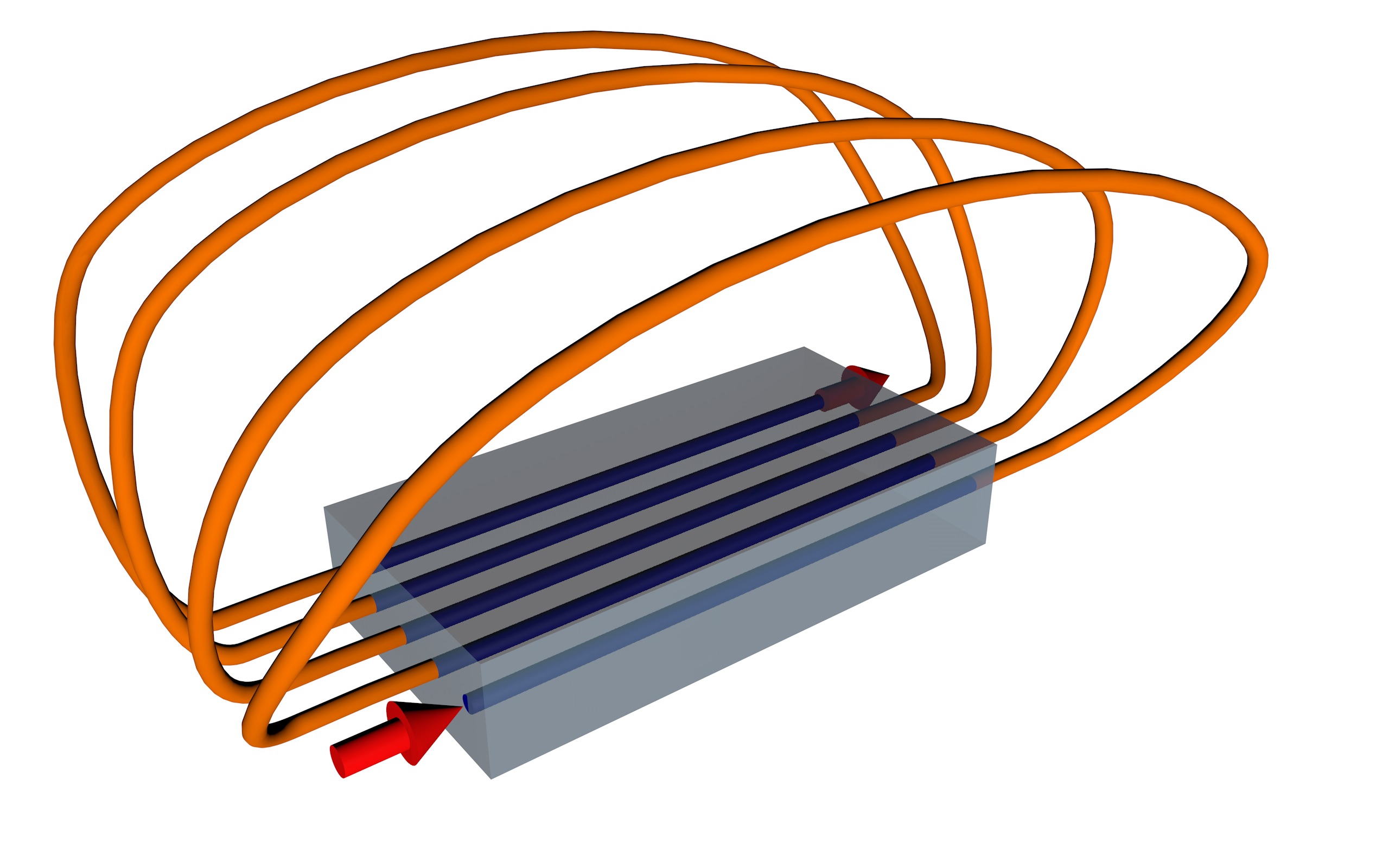}
\caption{{\bf Scheme of the proposed implementation.} Illustration of the chip for implementing the photonic simulator in Eq.~\eqref{chip}, where the arrows represent the input and output ports and the lines inside and outside the chip represent the waveguides and fiber connections respectively.}
\label{fig:fig1}
\end{figure}

In order to introduce our protocol, we start by considering the following first order linear and non-autonomous A-R differential equation
\begin{equation}
i\frac{dx(t)}{dt}=\beta(t)x(t)+\kappa^{-}(t)x(t-\tau)+\kappa^{+}(t)x(t+\tau),
\label{are}
\end{equation}
with $\kappa^+ (t)=\kappa^- (t+\tau)$, and boundary conditions $\kappa^- (0< t < \tau) = \kappa^+ ((N-1)\tau < t < N\tau)=0$. Associating the functions $x(t)$, $x(t\pm\tau)$ with the mode amplitude of monochromatic waves traversing the $j$-th and $j\pm 1$-th waveguides, $a_{j}(z)$ and $a_{j\pm 1}(z)$, of an array of $N$ evanescently coupled waveguides, each supporting a single mode and having a ``time" dependent propagation constant $\beta(z)$, we obtain a system governed by a set of $N$ differential equations  
\begin{equation}
i\frac{da_j}{dz}=\beta(z) a_j (z) + \kappa_{j, j+1} (z) a_{j+1} (z) + \kappa_{j,j-1} (z) a_{j-1} (z),
\label{chip}
\end{equation}
with $j\in [1,N]$. In the quantum optics regime, the dynamics of single photons traversing this type of devices is governed by a set of Heisenberg equations that are isomorphic to Eq.~\eqref{chip}. The only difference is that in the quantum case, the mode amplitude $a_{j}$ is replaced by the creation operators $a_{j}^{\dagger}$ \cite{Lai}. In order to make Eq.~\eqref{chip} isomorphic to Eq.~\eqref{are}, we must impose a continuity condition $a_j (0)=a_{j-1}(\tau)$ within the interval $j \in [2, N-1]$. Physically, this condition implies that the mode field $a_{j-1}$, at the propagation distance of $(j-1)\tau$ ($\tau$ is the length of the waveguides), has to be fed back into the input of the $j$-th waveguide. Furthermore, the finiteness of the waveguide array imposes the boundary conditions $\kappa_{1,0}=\kappa_{N,N+1}=0$. Additionally, we establish the aforementioned mapping of the independent variable $t$ into the spatial coordinate $z$. Therefore, our protocol requires the implementation of waveguide lattices endowed with input-output connections as illustrated in Fig.~\ref{fig:fig1}. We point out that the phase introduced via the fiber propagation can be traced and disregarded given that it can be made equal for all fiber connections, being therefore a global phase.

\begin{figure}[h!!]
\includegraphics[width=0.5\textwidth, trim= 0cm 5cm 0cm 5cm]{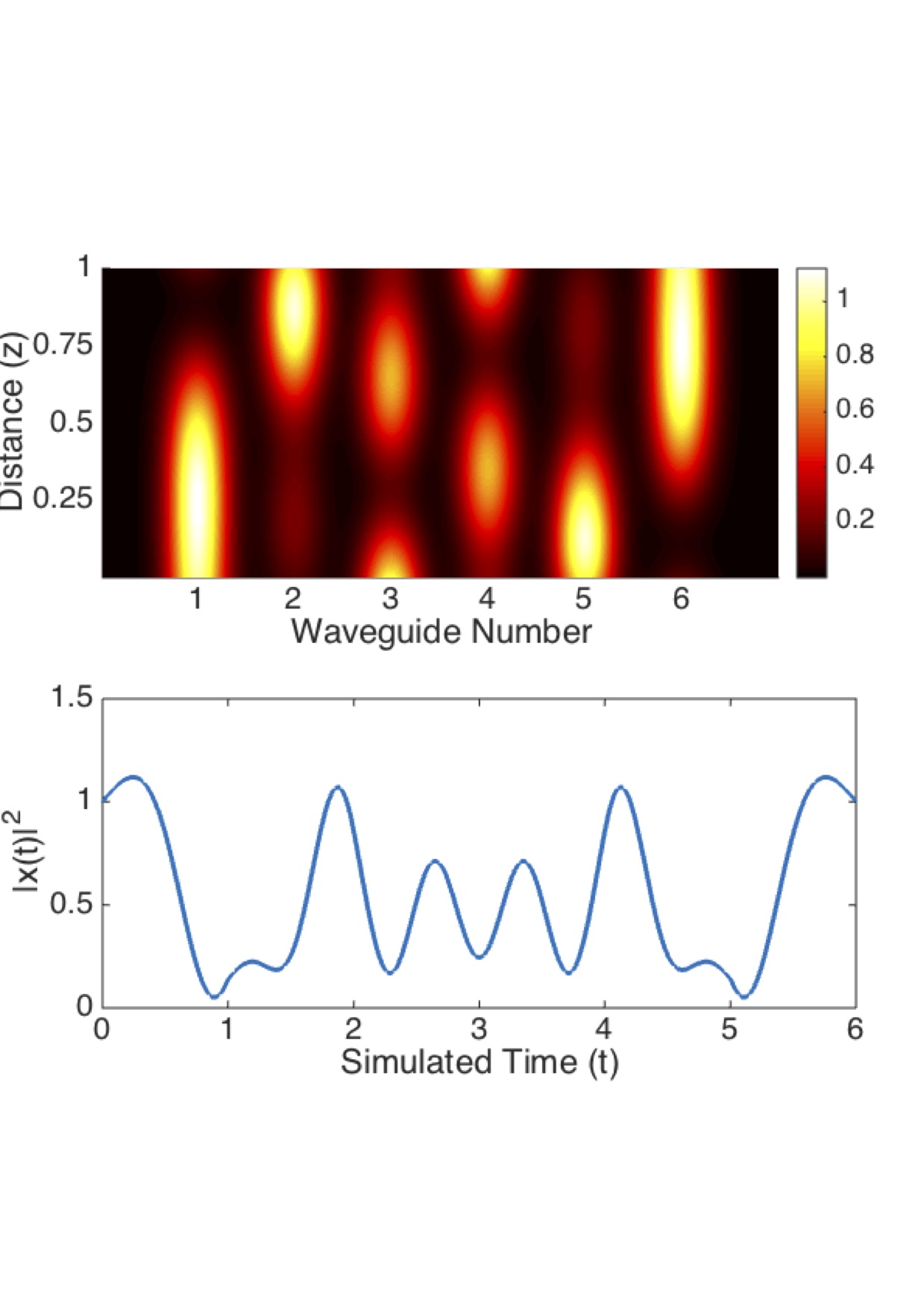}
\caption{{\bf Numerical simulations of the dynamics.} (a) Intensity evolution for an array having $N=6$ waveguides and constant lattice parameters $\beta~=~1$, $\kappa~=~\sqrt{\beta+N}$ and $\tau=1$. (b) Intensity of all the waveguides concatenated in a single curve, which represents the absolute square of the solution of Eq.~\eqref{chip}.}
\label{fig:fig2}
\end{figure}

The analogy we present here can be exploited with classical electromagnetic fields or single photons. The motivation for doing a quantum simulation is the possibility to embed the advanced-retarded dynamics in more general quantum protocols allowing the implementation of quantum feedback and feedforward. In the classical case, the initial condition $a_1 (0)$ is implemented by continuously injecting light into the system. This condition is crucial to establish the isomorphism between the light dynamics in the waveguide array and the solution of Eq.~\eqref{are}. As a result, the solution of Eq.~\eqref{are} is obtained in the stationary regime of our photonic system. Once the intensity is measured, the modulus square of the solution is obtained by merging the intensity evolution of each waveguide in a single variable. See Fig.~\ref{fig:fig2} for a demonstration of the potentiality of our protocol, in which we analyze the setup depicted in Fig.~\ref{fig:fig1}. The light dynamics occurring in such an array is governed by Eq.~\eqref{chip}. 

An interesting point to highlight here is the existence of a $z$ reversal symmetry in the simulation with respect to the central point of the evolution, $z_c=N\tau /2$, for constant lattice parameters. This relation holds for the modulus square of the solution, $a a^{*}(z_c +z)=a a^{*} (z_c -z)$. Consequently, after the system reaches the steady state, it simultaneously fulfills the periodic boundary condition, $a(0)=a(N\tau)$, where we have neglected a global phase factor. This property combined with the space-time analogy opens a possible framework for the study and implementation of Closed and Open Timelike Curve gates \cite{iqp1,iqp2,ctc1,ctc2,ctc3,ctc4,ctc5}.
 
\paragraph{Non-Autonomous Equations} A more general scenario arises when considering space-dependent parameters in Eq.~\eqref{chip}, which in the context of photonic lattices, means that the system becomes dynamic. The dynamic character is achieved with modulations of the refractive index of individual waveguides. We consider an implementable system conformed by a periodic variation $\beta(z)=\beta_{0}+\epsilon\cos(\omega z)$, where $\beta_{0}$ is a constant, $\epsilon$ is the modulation amplitude, and $\omega$ stands for the modulation frequency along $z$. For illustration purposes, we provide a numerical calculation of this particular photonics simulator in the Supplementary Fig. 1. Notice that in the context of advanced-retarded equations the time dependence allows us to encode non-autonomous equations, which are hard to compute in general.

\begin{figure*}[]
\includegraphics[width=15cm, trim= 0cm 0cm 0cm 0cm]{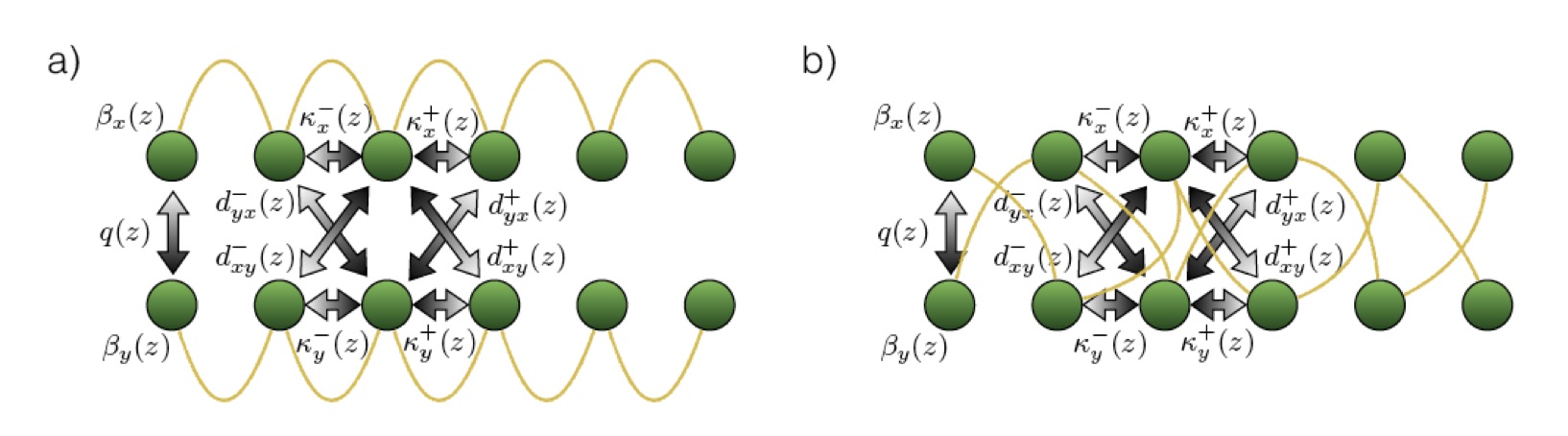}
\caption{{\bf Systems of equations.} Scheme of the chip in a perpendicular plane showing the input output connections and the parameters of the simulation. Here $\beta$ is the propagation constant, $q$ is the vertical coupling constant, $\kappa$ is the horizontal coupling constant and $d$ the diagonal coupling constant. a) The scheme in which each plane is associated with a component of the qubit simulates Eq.~\eqref{qubit}. b) The crossed links allow for a stronger temporal mixing of the qubit components in the derivative. This situation corresponds to the second example of Eq.~\eqref{qubit}.}
\label{fig:fig3}
\end{figure*}

\paragraph{Systems of Equations} We now turn our attention to demonstrate how our protocol can be extended to provide solutions of systems of A-R equations. To this end, we encode every unknown function in a waveguide array, and place all the involved arrays close to each other in such a way that light fields traversing the system can tunnel from array to array. In this manner, all functions are self-coupled and coupled to others, enriching the dynamics of the systems. In order to explain the operating principle of the protocol, we focus on the simplest case of two variables. Note that in this case, we can relate each array to a component of a qubit to be implemented in the dual-rail encoding with a single photon. As a first possibility, we can put together two lattices in which the feedback takes place as depicted in Fig.~\ref{fig:fig3}a. In this scenario, the light can hop to neighboring sites as well as to the sites of the adjacent array. This arrangement enables the time evolution simulation of a single qubit Hamiltonian combined with two terms corresponding to advanced and retarded couplings 
\begin{equation}
\label{qubit}
i|\dot{\psi}(t)\rangle=H(t) |\psi(t)\rangle + H^+ (t)|\psi(t+\tau)\rangle + H^- (t) |\psi(t-\tau)\rangle
\end{equation}
\begin{eqnarray}
\label{hamiltonian}
H(t)=\left( \begin{array}{cc} \beta_x (t)&q(t)\\q(t)&\beta_y (t) \end{array} \right), \qquad  H^+(t)=\left( \begin{array}{cc} \kappa^+_x (t)&d^+_{xy}(t)\\d^+_{yx}(t)&\kappa^+_y (t) \end{array} \right), \qquad H^-(t)=\left( \begin{array}{cc} \kappa^-_x (t)&d^-_{xy}(t)\\d^-_{yx}(t)&\kappa^-_y (t) \end{array} \right).
\end{eqnarray}

A second configuration arises by mixing the connectors as shown in Fig.~\ref{fig:fig3}b. In this case we flip the qubit in the advanced and retarded times. Even though the equation exhibits the same structure as Eq.~\eqref{qubit}, the forward and backward Hamiltonians, $H^+$ and $H^-$ are the same matrices given in Eq.~\eqref{hamiltonian} multiplied to the left by $\sigma_x$. Here $H(t)$, $H^+ (t)$ and $H^- (t)$ depend on the propagation constants, on the coupling between waveguides belonging to different arrays and the coupling between the guides of the same array. The vertical coupling is refereed to as $q$ while the transverse coupling constants are represented by $\kappa$ and the labels $(x,y)$ correspond to the plane where the arrays are located. Moreover, $d$ accounts for the diagonal coupling. In addition to the conditions for each array in \eqref{are}, the following consistency and boundary conditions are fulfilled, 
\begin{eqnarray}
\nonumber H^+ ((n-1)\tau < t < n\tau)= H^- (0< t < \tau)=0, \qquad  d^+_{yx}(t)=d^-_{xy}(t+\delta), \qquad d^+_{xy}(t)=d^-_{yx}(t+\delta).
\end{eqnarray}
In Supplementary Fig. 2 a numerical calculation of Eq.~\eqref{qubit} is provided.
Even though our formalism is based mainly on classical optics, one can emulate a quantum system (qubit) via the encoding of multiple A-R equations, as shown in this section. Moreover, one may also include quantum effects via the loading of the chip with genuinely quantum photonic states, e.g., two-photon or few-photon Fock states, which will introduce quantum effects as photon bunching, in combination with advanced-retarded physics. A thorough analysis of this last possibility is, however, outside of the scope of this article.

\paragraph{Multiple Delays} We next consider a variant of Eq.~\eqref{are} with multiple delays. This configuration arises when we allow each waveguide to couple to multiple neighbors and reordering the feedback connectors. The first non-trivial example is a two-time A-R equation, Eq.~\eqref{are2}. 
\begin{eqnarray}
\label{are2}
i\dot{x}(t)= \beta(t)x(t)+\kappa^+ (t) x(t+\tau)+\kappa^{++} (t) x(t+2\tau)+ \kappa^{-} (t) x(t-\tau) + \kappa^{--} (t) x(t-2 \tau).
\end{eqnarray}
Experimentally, the arrangement can be engineered by fabricating the waveguides in a zig-zag configuration. The resulting equation shares the structure of an A-R differential equation with additional feedback and feedforward terms. For this particular system the coupling coefficients are related as
\begin{equation}
k^{++}(t)=k^{--}(t+2\tau), \qquad k^{+}(t)=k^{-}(t+\tau),
\end{equation}
with the appropriate boundary conditions. See Supplementary Fig. 3 for a numerical simulation of Eq.~\eqref{are2}.

\paragraph{Higher Order Equations} One last generalization of the A-R simulator consists in introducing complex dynamical parameters. This can be achieved by combining our feedback technique with Bloch oscillator lattices \cite{bend1, bend2, bend3}. These types of arrays can be implemented by including a transverse ramping on the potential of the waveguides or by curving the waveguide arrays. Provided the evolution equations for the Bloch oscillator array, $i \dot{a_n}+ n\beta{z}a_{n} + \kappa^+ a_{n+1} + \kappa^- a_{n-1}=0$, and making the formal transformation $a_n=\tilde{a}_{n}(z)\exp(in\phi(z))$, with $\phi(z)=\int_{0}^z\beta(z')dz'$, one can show that it is formally equivalent to a system endowed with complex coefficients, $i\dot{\tilde{a}}_n + \exp(in\phi(z))\kappa^+ \tilde{a}_{n+1} + \exp(-in\phi(z))\kappa^-\tilde{a}_{n-1}$. The inclusion of arbitrary complex parameters could be used to enhance the versatility of the protocol. Furthermore, a complete control of the coupling constants would allow to simulate higher order equations via systems of first order equations.  

Even though the toolbox introduced here is valid for simulating diverse physical configurations, it can be generalized by an appropriate mathematical treatment. Let $M(t)$ be the matrix containing the information about the propagation constant and couplings among the waveguides defined by $i \dot{a}(t)=M(t) a(t)$, $F$ the matrix encoding the input-output connections, and $\alpha$ the initial state independent of the feedback and feedforward mechanism, such that $a(0)=F a(\tau) + \alpha$. Consider now that the dynamical equation is solvable in terms of the evolution operator $U(t)$, $a(t)=U(t) a(0)$. Our goal is to determine the complete initial condition $a(0)$ in terms of the independent initial condition $\alpha$, evolution operator $U(t)$ and input-output matrix $F$. The consistency relations at $a(\tau)$ impose a set of equations that $a(0)$ has to fulfill, $a(\tau)=U(\tau) a(0)$, 
\begin{eqnarray}
a(0)=F a(\tau) + \alpha \Rightarrow a(0)=\left( \mathbb{1}-FU(\tau) \right)^{-1} \alpha, \qquad a(t)=U(t)( \mathbb{1}-FU(\tau))^{-1}\alpha.
\end{eqnarray}   

Notice that this relation holds for any $\alpha$, allowing the input of quantum states superposed in more than one waveguide, and is also valid for different configurations of couplings $U$ and connections $F$, limited by the existence of the inverse of $(\mathbb{1}-F U(\tau))$. Moreover, we can think of different experimental conditions, in which the connections happen at distinct evolution times $\tau_i$, $a(0)=\sum F_i a(\tau_i)+\alpha$, resulting in $a(t)=U(t)( \mathbb{1}-\sum F_i U(\tau_i))^{-1}\alpha$.  

\subsection*{Implementation Analysis}

We analyze now the main limitations of the protocol regarding its interpretation and experimental realization. 

Regarding the quantum equation, although the probability is not normalized during the dynamics, it is normalized in the input and output points. Therefore, initialization and measurements retrieve the correct interpretation, even if the particle undergoes forward and backward jumps on time, an effect that could be useful for the simulation of absorbing potentials \cite{ile}. 

\begin{figure}[t]
\includegraphics[width=0.5 \textwidth, trim= 0cm 5cm 0cm 5cm]{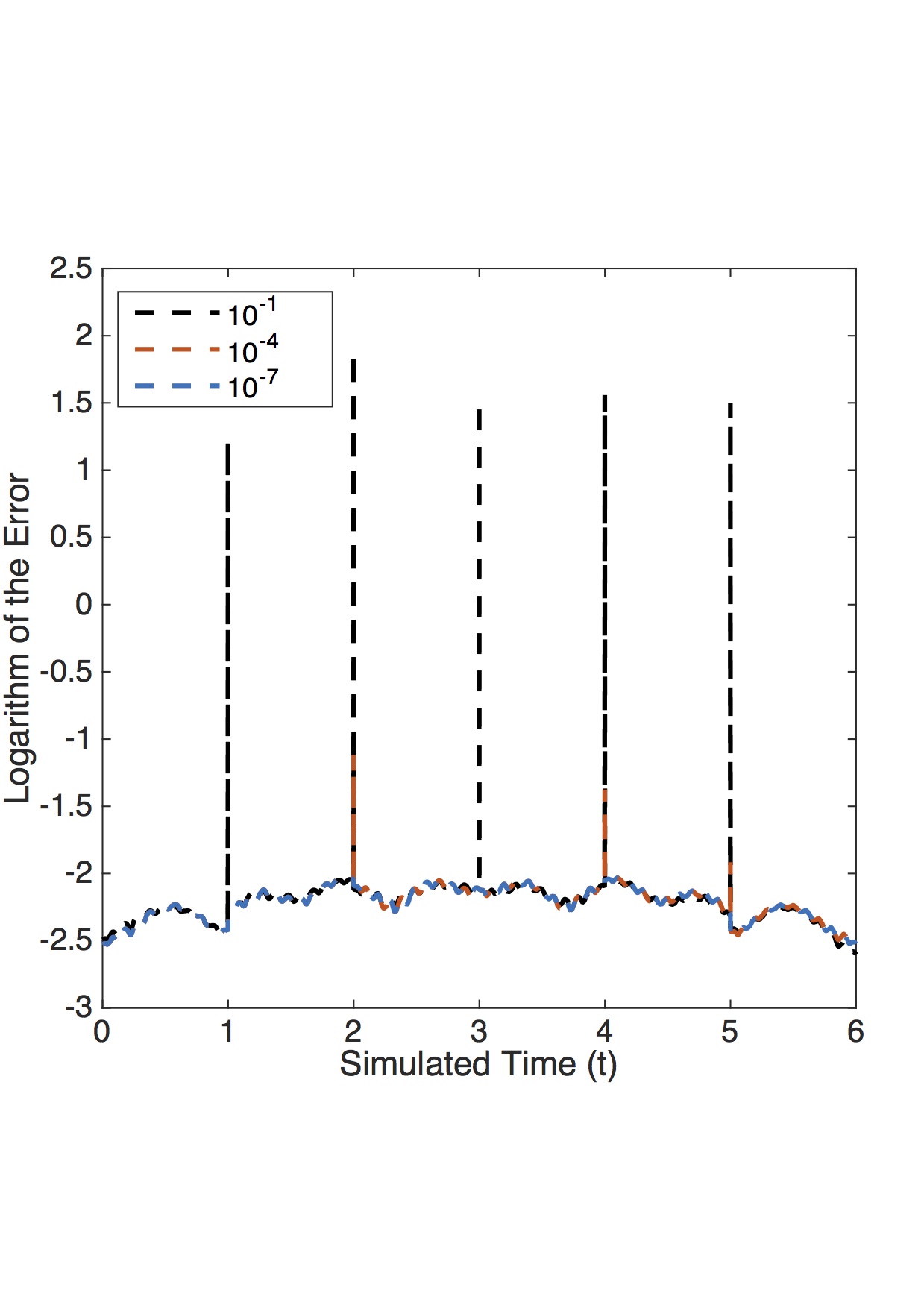}
\caption{{\bf Error analysis.} We depict the decimal logarithm of the error as a function of time for three runs of the simulation with different distance with respect to the stationary state. The fact that the effective interaction between photons is null makes possible the analogy between the stationary state solution and the accumulation of solutions for an initial excitation combined until the initial population has escaped from the output port. Therefore, the distance is calculated as the norm of the population that remains in the chip. The dynamical constants of the system are equivalents to the ones in Fig. \ref{fig:fig2}.}
\vspace{-0.5cm}
\label{fig:fig4}
\end{figure}

\paragraph*{Stationary State Proximity}
A natural source of errors is given by the waveguide losses damaging the quantum state in the time elapsed until the photon escapes from the chip. This time is related with the population in the solution, which is unknown before the experiment is realized. Identifying the relation between the resonances and the dynamical parameters could be helpful for estimating the population in the stationary state, and therefore the total experimental time for achieving the stationary state in the classical case. See Fig. \ref{fig:fig4} for a scheme of the error depending on the distance from the stationary state.   

\paragraph*{Experimental Errors}
We have to take into account that the losses introduced by the fiber connections will break the continuity condition allowing us to simulate Eq.~\eqref{are} in terms of Eq.~\eqref{chip}. The length and propagation constant of this fibers have to be tuned so that no phase is introduced in the evolution. Additionally, the space dependence of propagation and coupling constants is limited by the experimentally realizable functions. The degrees of freedom to be considered are the writing precision for modifying the propagation constants and the spatial path of each waveguide for modifying the coupling constants. 

\section*{Discussion}

In conclusion, we have developed a flexible but realistic toolbox for implementing advanced-retarded differential equations in integrated quantum photonics circuits. We have shown that our analogy enables the simulation of time dependent systems of advanced-retarded equations, which in the context of quantum information can be employed to realize feedback and feedforward driven dynamics. Therefore, we consider that our work enhances the versatility of quantum simulators in the abstract mathematical direction and in terms of applications for retrospective and prospective quantum memory.

\section*{Supplemental Information}
See the included Supplemental Information for supporting content.

\section*{Acknowledgements}
 We acknowledge support from Spanish MINECO/FEDER FIS2015-69983-P; Ram\'{o}n y Cajal Grant RYC-2012-11391; UPV/EHU UFI 11/55 and EHUA14/04; Basque Government BFI-2012-322 and IT986-16; a UPV/EHU postdoctoral fellowship; Deutsche Forschungsgemeinschaft (grants SZ 276/7-1, SZ 276/9-1, SZ 276/12-1, BL 574/13-1, GRK 2101/1); and the German Ministry for Science and Education (grant 03Z1HN31).

\section*{Author Contributions}
U. A.-R. and A. P.-L. made the numerical simulations while U. A.-R., A. P.-L., I. L. E., M. G., M. S., L. L., A. S., and E. S. developed the protocol and wrote the manuscript.

\section*{Additional information}
The authors declare no competing financial interests.

\clearpage

\part*{Supplementary Information}
This document provides supplementary information to Advanced-Retarded Differential Equations in Quantum Photonic Systems. In particular, we provide examples of the genuine physical situation developed in the manuscript each of them associated with a given A.-R. equation, which show the versatility of our analog photonics simulator.

\begin{figure*}[htbp]
\centering
\includegraphics[width=\linewidth]{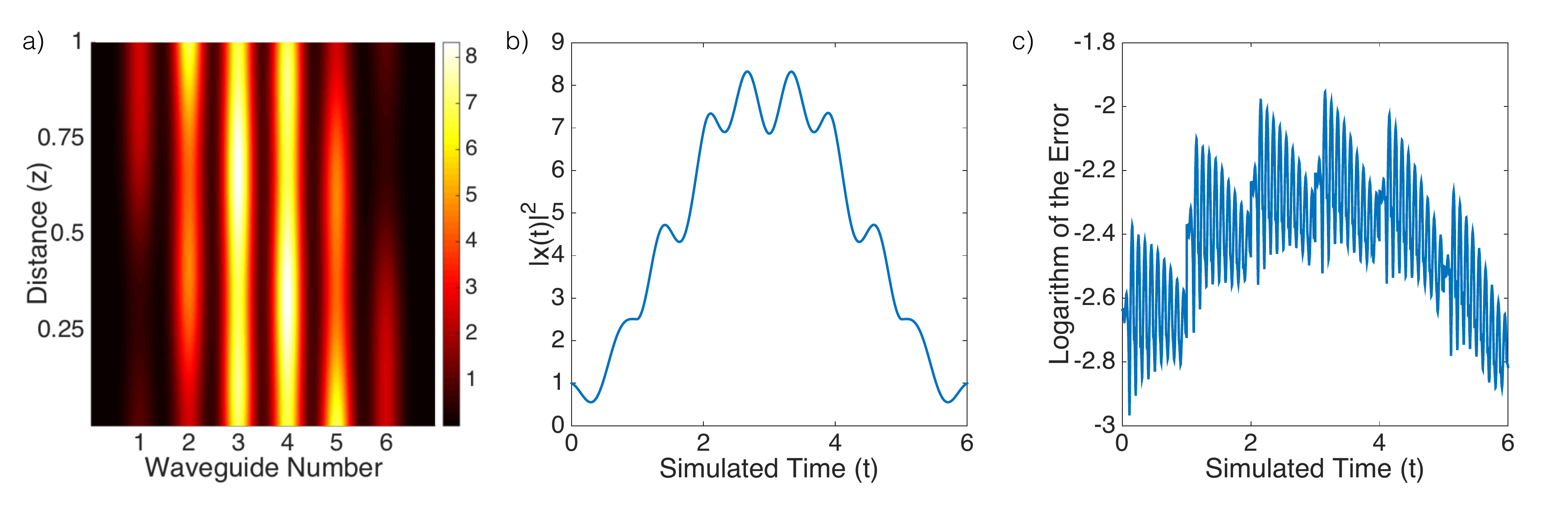}
\caption{Supplementary Figure 1. Numerical Simulation of Eq.~$(1)$ with $N=6$, $\beta=1$, $\epsilon=1$, $\kappa=\sqrt{7}$, $\omega=2$.  (a) Intensity in the stationary state in the waveguides array. (b) Modulus square of the solution as a function of time. (c) Decimal Logarithm of the error of the simulation with respect to the solution of the A-R equation.}
\label{fig:s1}
\end{figure*}

\section*{Non-Autonomous Equations}
Consider the oscillatory time dependence of the propagation constant, $\beta$, for the single variable A-R equation, Eq.~$(1)$ in the manuscript. We numerically simulate this system for a lattice of $N=6$ waveguides, $\beta=1$, $\kappa=\sqrt{7}$, $\epsilon=1$ and $\omega=2$, see Fig. \ref{fig:s1}.  We have selected this non-autonomous advanced-retarded equation to show the existence resonant solutions, in which a high amount of light gets trapped in the chip. Notice that although a high population is achieved in the stationary state the theoretical error is still small. 

\begin{figure*}[htbp]
\centering
\includegraphics[width=\linewidth]{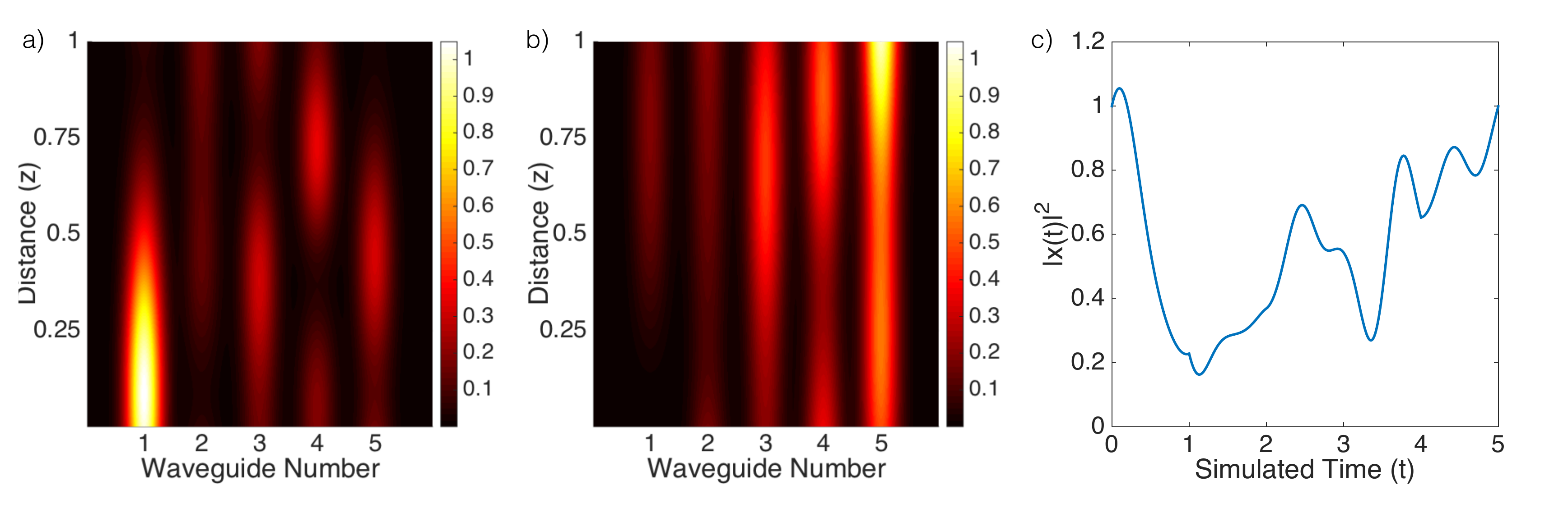}
\caption{Supplementary Figure 2. Numerical Simulation of Eq.~$(3)$ with $N=5$, $\beta_x=1$, $\beta_y=2$, $\kappa_x=3$, $\kappa_y=1$, $q=1$, $d=1$, $\tau=1$ for the initial state $|\psi(0)\rangle=|0\rangle.$  (a) Waveguides intensity in the $x$ plane corresponding to the first component of the qubit. (b) Waveguides intensity in the $y$ plane corresponding to the second component of the qubit. (c) Modulus square of the quantum state as a function of time.}
\label{fig:s2}
\end{figure*}

\section*{Systems of Equations}
In Fig. \ref{fig:s2} we show a numerical simulation of the array proposed to simulate systems of A-R equations given by Eq.~$(3)$. The dynamical parameters are the following ones, $N=5$, $\beta_x=1$, $\beta_y=2$, $\kappa_x=3$, $\kappa_y=1$, $q=1$, $d=1$ and $\tau=1$ for an initial state $|\psi(0)\rangle = |0\rangle$. The theoretical error is smaller than $10^{-2}$ for the complete time evolution. We have selected this parameters to show that highly asymmetric solutions are also possible for time independent equations even with the limitation induced by the physical constrains in the coupling constants $\kappa$.

\begin{figure*}[htbp]
\centering
\includegraphics[width=\linewidth]{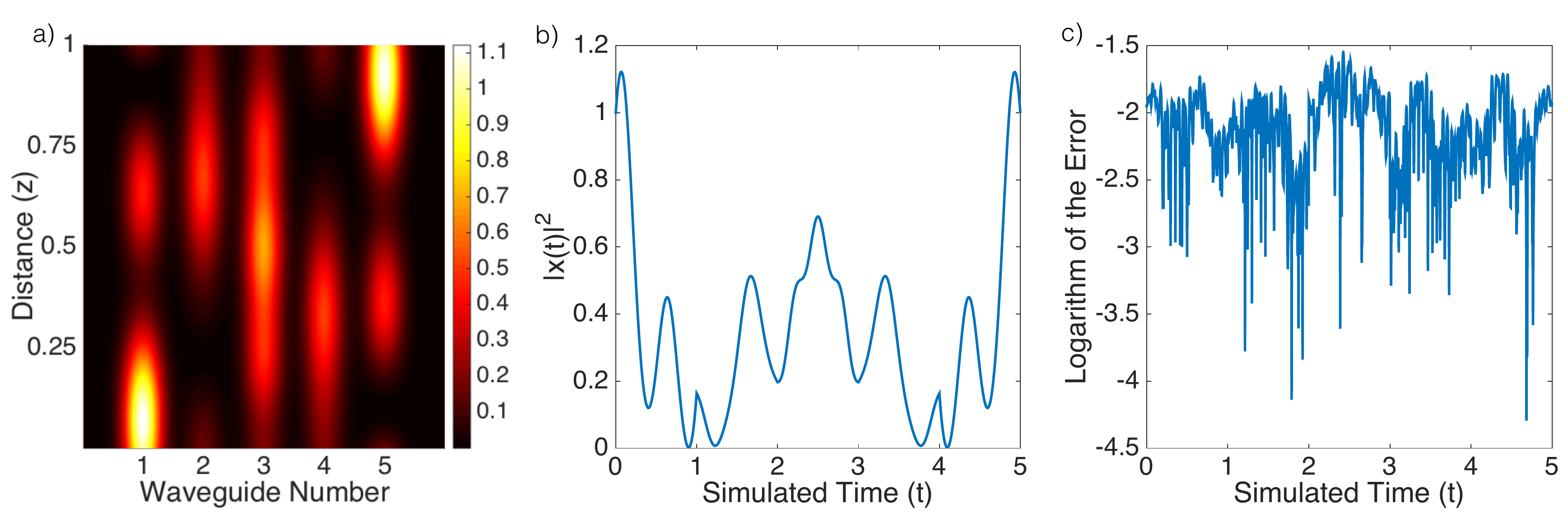}
\caption{Supplementary Figure 3. Numerical simulation of Eq. $(6)$ with $N=5$, $\beta=1$, $\kappa=5$ and $\tau=1$. (a) Intensity in the stationary state in the waveguides array. (b) Modulus square of the solution as a function of time. (c) Decimal Logarithm of the error of the simulation with respect to the solution of the A-R equation.}
\label{fig:s3}
\end{figure*}

\section*{Multiple Delays}
Eq.~$(6)$ in the manuscript describes the evolution of a system driven by two feedback and two forward terms. See Fig. \ref{fig:s3} for a numerical simulation of this equation with $N=5$, $\beta=1$, $\kappa=5$ and $\tau=1$.

\end{document}